\documentstyle[12pt]{article}
\renewenvironment{figure}[7]{ }{}
\begin{document}
\begin{titlepage}
\begin{center}
May 1994\hfill    SU-ITP-94-15 \\
               \hfill    RU-94-37 \\
               \hfill    hep-ph/9405313
\vskip .2in
{\large \bf
Yukawa Unification: The Good, The Bad and The
Ugly}\footnote{Presented at the Second IFT Workshop on Yukawa
Couplings and the Origins of Mass, 11-13 February 1994, Gainesville,
Florida.}
\vskip .3in
Riccardo Rattazzi\footnote{E-mail:
rattazzi@physics.rutgers.edu}\\[.12in]
{\em Department of Physics and Astronomy\\
     Rutgers University\\
     Piscataway, NJ  08855}
\vskip 13pt
Uri Sarid\footnote{E-mail:
sarid@squirrel.stanford.edu}~(presenter)\\[.12in]
{\em Physics Department\\
     Stanford University\\
     Stanford, California 94305}
\vskip 13pt
Lawrence J. Hall\footnote{
E-mail: hall\_lj@theorm.lbl.gov}\\[.12in]
{\em Theoretical Physics Group, 50A/3115\\
      Lawrence Berkeley Laboratory\\
     1 Cyclotron Road\\
     Berkeley, California 94720}
\end{center}
\vskip .2in
\begin{abstract}
We analyze some consequences of grand unification of the
third-generation Yukawa couplings, in the context of the minimal
supersymmetric standard model. We address two issues: the prediction
of the top quark mass, and the generation of the top-bottom mass
hierarchy through a hierarchy of Higgs vacuum expectation values.
The top mass is strongly dependent on a certain ratio of
superpartner masses. And the VEV hierarchy always entails some
tuning of the GUT-scale parameters. We study the RG equations and
their semi-analytic solutions, which exhibit several interesting
features, such as a focusing effect for a large Yukawa coupling
in the limit of certain symmetries and a correlation between the
$A$ terms (which contribute to $b\rightarrow s\gamma$) and the
gaugino
masses. This study shows
that non-universal soft-SUSY-breaking masses are favored (in
particular for splitting the Higgs doublets via D-terms and for
allowing more natural scenarios of symmetry breaking), and hints at
features desired in Yukawa-unified models. Several phenomenological
implications are also revealed.
\medskip
\end{abstract}
\end{titlepage}


\def\figone{The dependence of the {\it low-energy} values
$\lambda_{t,b,\tau}$ and of the ratio $R = \lambda_b/\lambda_\tau$
on the initial condition $\lambda_{t,b,\tau}^G=\lambda_G$, without
any
threshold corrections. The allowed range of $R_{\rm expt}$ is shown
shaded, and the minimal value of $\lambda_G$ allowed by this range
in the absence of any corrections is indicated by the solid dot;
lower values of $\lambda_G$ require finite, negative $\delta m_b$
(see the text). The corresponding value of $\lambda_t$ is marked by
the shaded dot. The vertical scale on the right indicates the
approximate tree-level top mass $\sim 174 \lambda_t\rm\,GeV$ which
would result from the values of $\lambda_t$ on the left vertical
scale; for example, the shaded dot predicts a heavy top, above 170
GeV or so.}

\def\figtwo{The leading (finite) 1-loop MSSM corrections to the
bottom quark mass, namely $\delta m_b$.}

\def\figthree{Our predictions \cite{ref:us}\  for
the pole mass of the top quark, without superheavy corrections and
using two qualitatively-different
superpartner spectra, specifically $m_{\rm higgsino} \sim \mu = 100
\GeV$, $m_{\rm gluino} = 300\GeV$,
$m_{\rm wino} = 100\GeV$, $m_{\rm squark} = m_{\rm slepton} =
1000\GeV$ and $m_A =
1000\GeV$ for which the $\delta m_b$ corrections are small, and
$m_{\rm higgsino} \sim \mu = 250
\GeV$, $m_{\rm gluino} = 300\GeV$,
$m_{\rm wino} = 100\GeV$, $m_{\rm squark} = m_{\rm slepton} =
400\GeV$ and $m_A =
400\GeV$, for which $|\delta m_b/m_b|$ $\,\sim 0.25$. The upper or
lower
horizontal axes should be used for these two spectra, respectively.
The ``cloud'' indicates the region where the theory becomes
nonperturbative at the GUT scale. Also shown are the estimated
allowed mass ranges for the running parameter $m_b$ as extracted in
our previous work \cite{ref:us}.}

\def\GeV{{\rm\,GeV}}
\def\pha{\phantom{m_U\,}}
\def\phb{\phantom{B\,}}
\def\half{{\textstyle{1\over2}}}
\def\fra#1#2{{\textstyle{#1\over#2}}}
\def\roughly#1{\,\,{\raise.3ex\hbox{$#1$\kern-.75em
\lower1ex\hbox{$\sim$}}}\,\,}

\section{Introduction}

There is strong evidence to suggest that the three gauge couplings
of the
strong, electromagnetic and weak interactions are unified at a high
energy
scale in a single gauge interaction based on a simple group, such as
SU(5) or SO(10), as long as the desert below the unification scale
is described by the minimal supersymmetric extension of the standard
model (the MSSM). Furthermore,  the combination of supersymmetry
(SUSY) and grand
unification yields models with numerous attractive features: the
embedding of the standard-model matter multiplets into a few
irreducible representations of the GUT group, the technically
natural preservation of a
hierarchy between the weak and GUT scales, a longer proton lifetime
to allow
agreement with current experimental lower bounds, the correct
prediction of
the ratio of $b$ quark to $\tau$ lepton masses, simple ans\"atze for
the
remaining fermion masses, and a picturesque scenario for radiative
electroweak
symmetry breaking. We have chosen, therefore, to look  beyond the
gauge
unification prediction of the weak mixing angle and examine the
unification of
third-family Yukawa couplings \cite{ref:yukuni}, for the most part
within its natural context of SO(10) unification. By
Yukawa couplings we mean the couplings of the top, bottom and tau to
the Higgs
doublets which generate their masses when electroweak symmetry is
broken.
The third family is singled out because of its
relatively large
Yukawa couplings: it seems reasonable to suppose that they arise at
tree-level
from the simplest interactions, while the masses and mixings of the
other
generations require more complex, perhaps also higher-order and
certainly very
model-dependent structures. Our study also applies more generally to
scenarios
in which these Yukawa couplings are unified but not in the context
of an
SO(10) GUT, or even models in which the Yukawas are only comparable
near the
GUT or Planck scales. We begin by considering the most
immediate
prediction of this Yukawa unification, namely the top quark mass
\cite{ref:us}. However, we
are quickly led to consider in some depth the more general question
of how the
top-bottom mass hierarchy could be generated in the MSSM, and how
this
hierarchy depends on the initial conditions of the renormalization
group (RG)
evolution at the GUT scale \cite{ref:us2}. We will conclude with a
discussion of how natural
(or unnatural!) such a hierarchy seems in this context, what its
other
phenomenological predictions might be, and how one could hope to
improve the
theoretical picture or obtain experimental corroboration.

One consequence of Yukawa unification is immediate, and independent
of other
assumptions except for the qualitative nature of the RG evolution
equations of
the MSSM. Since the Yukawa couplings of the top and bottom quarks
(and the $\tau$ lepton) are always
comparable, the large ratio of the top mass versus the bottom (or
tau) mass
must be due to a large ratio of the Higgs vacuum expectation values
(VEVs)
which give rise to their masses. Namely, since the up-type and
down-type
matter fermion masses arise from couplings to the up-type and
down-type Higgs
multiplets ($H_{U,D}$), respectively, the large ratio
$m_t/m_{b,\tau} =
(\lambda_t v_U)/(\lambda_{b,\tau} v_D)$ is not a consequence of a
large ratio
of Yukawas $\lambda_t/\lambda_{b,\tau}$ but rather of large $v_U/v_D
\equiv
\tan\beta$. Thus Yukawa unification generically implies $\tan\beta
\sim {\cal
O}(50)$. Further assumptions are necessary to make any precise
predictions. We will assume the following three throughout most of
this
work,
although we will point out those conclusions which are more general:
\begin{enumerate}
\item[(I)] The masses of the third generation, $m_t$, $m_b$ and
$m_\tau$, originate
from renormalizable Yukawa couplings of the form
$\underline{16}_3\,{\cal O}\,\underline{16}_3$ in a supersymmetric
GUT with a gauge group containing (the conventional) SO(10);
$\underline{16}_3$ denotes the 16-dimensional spinor representation
of SO(10)
containing the third-generation standard-model fermions (plus the
right-handed
neutrino which we assume to be superheavy) and their superpartners.
\item[(II)] The evolution of the gauge and Yukawa couplings in the
effective theory beneath the SO(10) breaking scale
is described by the RG equations of the MSSM.
\item[(III)] The two Higgs doublets lie
predominantly in a single irreducible multiplet of SO(10).
\end{enumerate}
The first and third assumptions serve to define what we mean by
Yukawa
unification, while the second allows us to relate this GUT-scale
unification
to weak-scale observables. From (I) and (III) it follows that the
third-generation Yukawas must arise from either a
$\underline{16}_3\,\underline{10}_H\,\underline{16}_3$ or a
$\underline{16}_3\,\underline{126}_H\,\underline{16}_3$ interaction
with an
SO(10) Higgs multiplet. The latter leads to the boundary conditions
$3 \lambda_t^G = 3 \lambda_b^G = \lambda_\tau^G \equiv \lambda_G$ at
the GUT
scale, but the resulting ratio of $m_b/m_\tau$ at low energies is
far too low
to be consistent with experiment (at least within the perturbative
regime, and
unless very large threshold corrections to the $b$ mass arise at low
energies \cite{ref:us}). Thus we are restricted to using the
$\underline{10}_H$, and hence
the boundary condition
\begin{equation}
\lambda_t^G = \lambda_b^G = \lambda_\tau^G \equiv \lambda_G\,.
\label{eq:bc}
\end{equation}

With this boundary condition, and using the unification of gauge
couplings to
fix the unification scale and the gauge coupling at that scale, we
can now
evolve the Yukawa couplings down to the weak scale for any given
value of
$\lambda_G$. The idea is that the three observable masses $m_t$,
$m_b$ and
$m_\tau$ are functions of the four parameters $\lambda_t$,
$\lambda_b$,
$\lambda_\tau$ and $\tan\beta$, and $\lambda_{t,b,\tau}$ are in turn
determined by the
unification in terms of $\lambda_G$ and the GUT scale $M_G$. Since
the latter
is already known from gauge unification, we are left with three
observable
masses as functions of only two parameters, $\lambda_G$ and
$\tan\beta$. Thus
we use two observables, $m_b$ and $m_\tau$ to fix $\lambda_G$ and
$\tan\beta$,
and thereby predict the third observable $m_t$. A detailed analysis
of the RG
evolution, and the consequent predictions, has already been
presented \cite{ref:us}. The
results, namely the values of $\lambda_{t,b,\tau}$ and the ratio $R
\equiv
m_b/m_\tau = \lambda_b/\lambda_\tau$ all at the weak scale, are
plotted in
Fig.~1 as functions of $\lambda_G$. (These curves actually use
2-loop RG
evolution, but at this point the difference between 1- and 2-loop
equations is
not important. For the final predictions of $m_t$ we use the full
2-loop evolution
and 1-loop matching conditions.) Evidently, larger $\lambda_G$
values correspond
to a heavy top and to a smaller $R$ ratio. The experimental value
$R_{\rm
expt}$ is found \cite{ref:us}\ from the QCD sum rules value and is
evolved to the weak scale
using 2-loop QCD running. We find, allowing for $\alpha_s$ to vary
between
roughly 0.11 and 0.12, the shaded range shown in Fig.~1. Thus, in
the absence
of any large corrections in the matching between the $R$ evolved
down from the
GUT scale and the $R_{\rm expt}$ in the standard model, we find
$\lambda_G >
0.75$, which implies a heavy top.

\begin{figure}{14}{0 160 700 500}{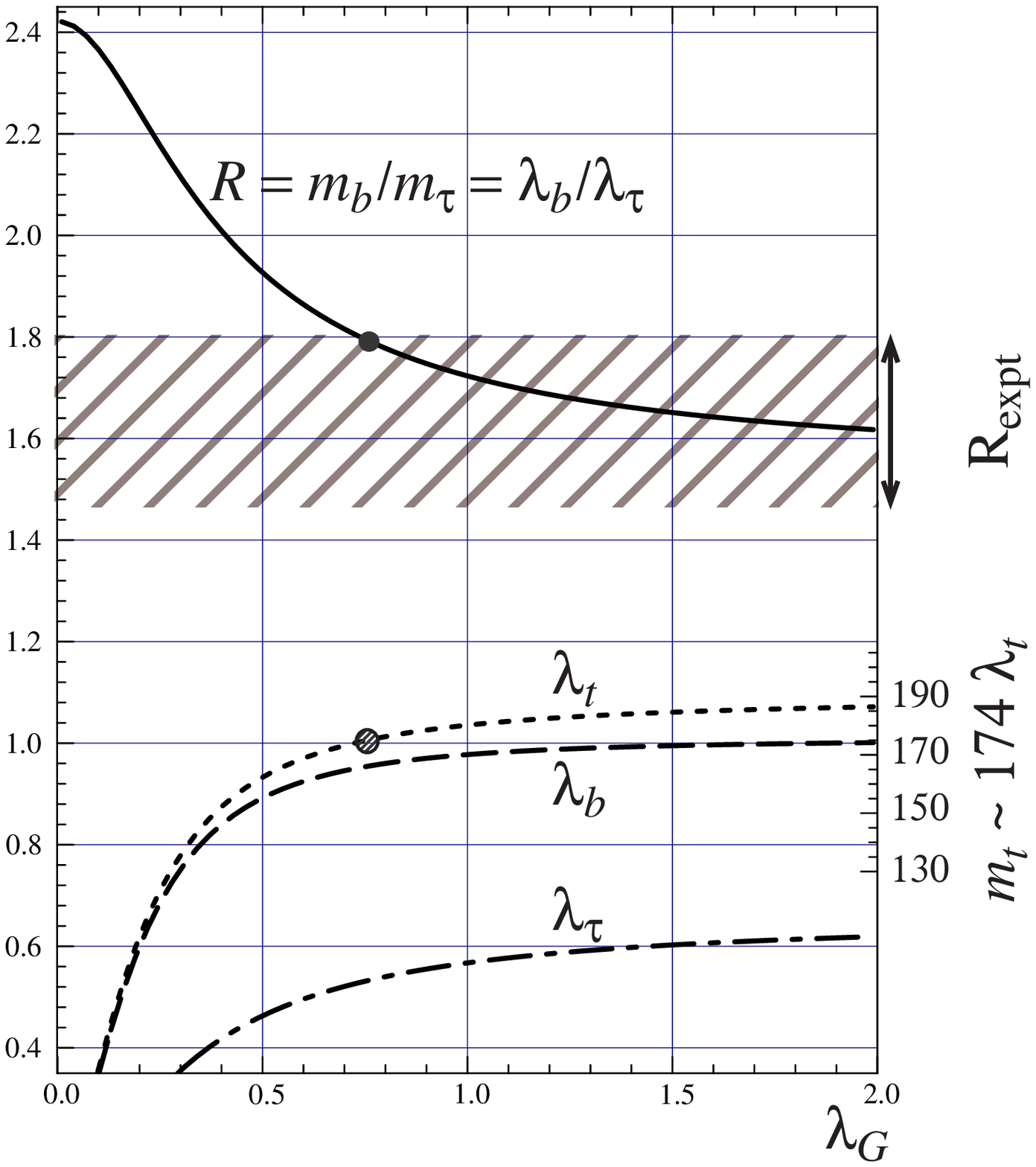}{1}{\figone}{3}{3}
\end{figure}

\section{The top and bottom masses}

For a precise prediction, 2-loop RG equations must be used along
with 1-loop
matching functions. These matching functions include logarithmic
corrections
from infinite counterterms as well as nonlogarithmic contributions
from finite
graphs. The former are given elsewhere \cite{ref:us}; they are
generally quite small, and
invariably increase the top mass as the superpartner masses
increase. The
latter are more interesting, since they can be very large
\cite{ref:us,ref:tb}. The dominant
corrections arise from the graphs of Fig.~2, which match the value
of the $b$
mass as evolved down from the GUT scale to the value in the
low-energy theory.
Typically the gluino graph dominates, yielding a corrected value
$m_b =
\lambda_b v_D + \delta m_b$ where $v_D = 174\rm\,GeV$,
\begin{equation}
{\delta m_b\over m_b} =  {8\over3}\, g_3^2\,
{\tan\beta\over16\pi^2}\,
{m_{\tilde{g}}\mu\over m_{\rm eff}^2}\,,
\label{eq:delmb}
\end{equation}
and $m_{\tilde{g}}$ is the gluino mass while $m_{\rm eff}$ is the
mass of the heaviest superpartner
in the loop (more exact expressions may be found in our previous
work). The
important observation here is that the bottom mass gets a large
contribution
from the up-type Higgs at 1-loop order, whereas its tree-level mass
was small
due to the small VEV of the down-type Higgs. In the usual scenario
with small
$\tan\beta$, the bottom was light because its Yukawa coupling was
small, or in
other words because it was protected by an approximate chiral
symmetry. Thus
any higher-order corrections would also be suppressed by this
approximate
symmetry. In the large $\tan\beta$ scenario these corrections are
not
suppressed: there is an enhancement of $v_U/v_D = \tan\beta$, which
overcomes
the usual $g_3^2/16\pi^2$ loop factor to give a correction of order
1 to the
$b$ mass---at least if $m_{\tilde{g}}\mu \sim m_{\rm eff}^2$.
Phenomenologically, the result is that $m_b$ cannot be predicted
with any
certainty unless we know something about the superspectrum, and if
$m_b$ is
uncertain then so is the top mass prediction.

\begin{figure}{14}{0 440 700 680}{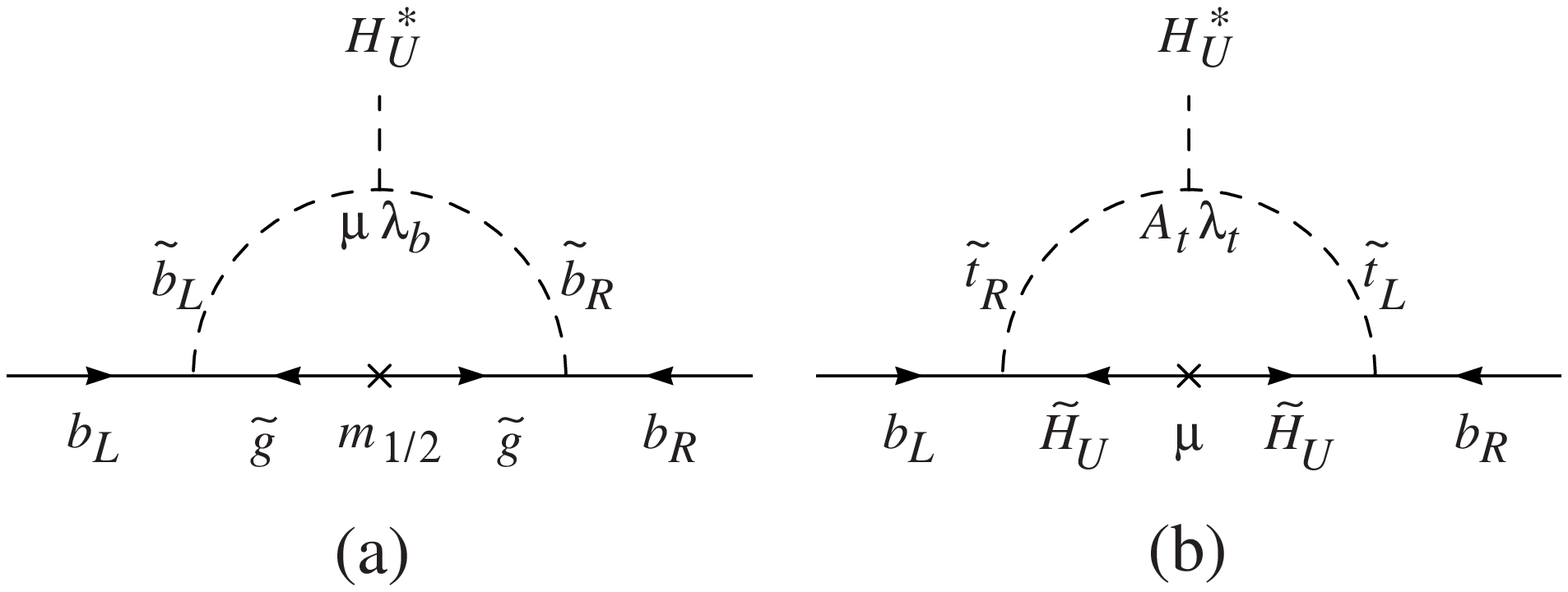}{2}{\figtwo}
{4}{2}
\end{figure}

In Fig.~3 we show the top pole mass prediction, now to full 2-loop
order, as a
function of the $\overline{\rm MS}$ running parameter $m_b(m_b)$ in
two cases.
The top curves and the higher horizontal axis correspond to a
hierarchical
spectrum, in which the squarks are heavy whereas the $\mu$ parameter
and
gaugino masses are light. Then the corrections $\delta m_b$ are
small and the
top mass is predicted to be above 180 GeV or so. The bottom curves
and the
lower horizontal axis correspond to a roughly degenerate spectrum in
which the
corrections to the $b$ mass are ${\cal O}(25\%)$ and negative (i.e.
$R$ should
be lowered by 25\% in Fig.~1 before matching to the experimental
value). Now
the top can be significantly lighter.
In fact, this last argument can be turned around: if the threshold
corrections
are too large in magnitude and negative then the top mass prediction
will be
below the experimental lower bound, while if they are large and
positive then
no value of $\lambda_G$ will allow agreement with $R_{\rm expt}$. We
thus find
the following bounds on the superspectrum: if $\lambda_G$ is allowed
to vary,
then
\begin{equation}
-0.37 < {m_{\tilde{g}}\mu\over m_{\rm eff}^2} < 0.08\,,
\label{eq:muglimits}
\end{equation}
whereas for fixed $\lambda_G$ the appropriate limits can be read
from Fig.~1.

\begin{figure}{15}{0 180 650 600}{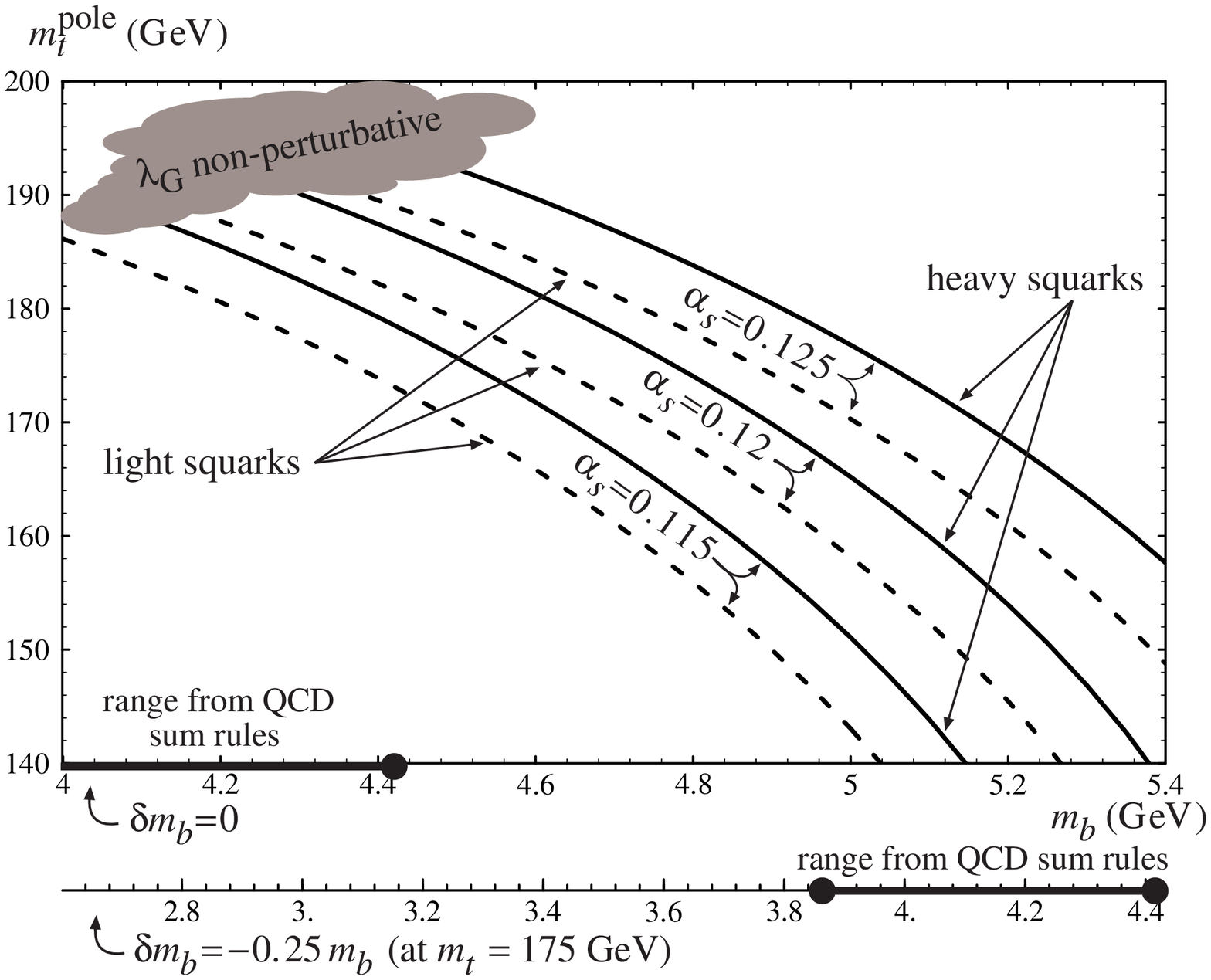}{3}{\figthree}
{5}{4}
\end{figure}

\section{Radiative bottom decay (I)}

Do we have any experimental information about these corrections? Of
course we
have no direct evidence of any of the superpartners, but we can
appeal to
their indirect appearance in loop diagrams. Consider again the
diagrams of Fig.~2,
but with one of the $b$ quarks replaced by a strange quark using a
flavor-changing vertex, and with a photon attached in all possible
ways. We
see that the same processes will lead to a contribution to the rare
decay
$b\rightarrow s\gamma$, and with a similar enhancement of ${\cal
O}(\tan\beta)$ over the usual MSSM scenario \cite{ref:us,ref:bsg}.
The dependence of these diagrams
on the superpartner masses is somewhat similar to that of $\delta
m_b$, except
that (a) now the operator is of higher dimension and so is
suppressed by the
mass of the heaviest superpartner, and (b) typically the
higgsino-mediated
diagram dominates. If the parameters appearing in this Higgsino
diagram,
namely $\mu$, $A$ (the trilinear soft SUSY-breaking parameter) and
the squark
masses, are all comparable and of order the Z mass, then this
diagram gives a
contribution to the amplitude for $b\rightarrow s\gamma$ many times
bigger
than the standard model or the usual MSSM amplitudes, and is clearly
ruled out
by the CLEO limit \cite{ref:cleo}. To restore agreement with
experiment, either the overall
superpartner mass scale must be raised far above the electroweak
scale, or
else $\mu$ or $A$ or both must be suppressed relative to the squark
mass in
the loop. More quantitatively, we find that either the masses must
all be
raised to at least ${\cal O}({\rm TeV})$, or else if both $\mu$ and
$A$ are
near the Z mass then the squarks must be above either $\sim 400$ GeV
or $\sim
700$ GeV, depending on whether these diagrams interfere
destructively or
constructively with the 2-Higgs standard model amplitudes. In any
case,
however, these restrictions do not yet tell us anything about the
combination
$m_{\tilde{g}}\mu/ m_{\rm eff}^2$ which appears in $\delta m_b$; the
link
between these two will be forged below, when we study the evolution
of the
entire set of MSSM parameters.

\section{Electroweak symmetry breaking}

We have assumed in the above that the top-bottom mass hierarchy
would arise
from a hierarchy of VEVs in the Higgs spectrum, namely $v_U/v_D
\equiv
\tan\beta \sim {\cal O}(50)$.  We now consider how such a hierarchy
could be
generated in the MSSM. Clearly this question is of interest for any
model in
which the Yukawas themselves do not supply a sufficient hierarchy to
explain
the large ratio of third-generation quark masses, not just for the
equal-Yukawas case
to which we have specialized. In studying this question, however, we
will
pay attention to which spectra are favored by large $\tan\beta$
scenarios, and
whether with such spectra and our unification assumptions we can pin
down the top mass prediction.

The scalar potential of the neutral Higgs bosons which leads to
electroweak
symmetry breaking is given by
\begin{equation}
V_0 = m_U^2 |H_U|^2 + m_D^2 |H_D|^2 + \mu B (H_U H_D + {\rm h.c.})
+ ({\rm quartic\,\,terms})
\label{eq:scapot}
\end{equation}
where $m_{U,D}^2 = \mu_{U,D}^2 + \mu^2$ contain the soft-breaking
masses and
the $\mu$ parameter in the superpotential, $B$ is a soft-breaking
mass
parameter, and the quartic terms arise from D-terms and so are given
by gauge
couplings. We will evolve these parameters from the GUT to the weak
scale
using the 1-loop RG equations of the MSSM, stopping the evolution at
some
typical scale (of order the squark masses) which minimizes the
effects of
higher-order corrections. The subscript 0 indicates that we will
restrict our
attention to this (RG-improved) tree-level potential rather than
calculate the
full 1-loop effective potential or, better yet, explicitly integrate
out
massive particles and consider full 1-loop matching conditions. We
expect \cite{ref:us2}\ that
our qualitative discussion of the radiative symmetry breaking will
not be
jeopardized by this simplification. That is, a more complete
calculation will
change the numerical values of the GUT parameters needed for
correctly
breaking the symmetry, but will not significantly alter the size of
the
domains in parameter space where such breaking is achieved. The
conditions for
this breaking are well-known:
\begin{equation}
m_U^2 + m_D^2 \geq 2 |\mu B|
\label{eq:bndbel}
\end{equation}
ensures that the potential is bounded from below, and
\begin{equation}
m_U^2 m_D^2 < \mu^2 B^2
\label{eq:ewsb}
\end{equation}
guarantees the existence of a minimum away from the origin and so
breaks the
symmetry. In practice, since $|\mu B|$ will always be much less than
or at
most comparable to $|m_U^2|$ and $|m_D^2|$, we can reduce these
requirements
to $m_A^2 = m_U^2 + m_D^2 > 0$ (using the expression for the
pseudoscalar
Higgs mass) and $m_U^2 < 0$ (noting that large $\tan\beta$ means
that the
up-type Higgs gets the large VEV).

In the usual---and very attractive---scenario of radiative breaking,
the two
mass parameters start out at the GUT scale with a universal positive
value:
$m_U^2 = m_D^2 = M_H^2 + \mu^2$, where $M_H$ is the soft-breaking
mass of the
$\underline{10}_H$ of Higgs in SO(10), or of
the $\underline{5}_H$ and $\underline{\overline{5}}_H$ in SU(5).
Thus the
symmetry is not broken at that scale. However, in the RG evolution
to the
electroweak scale, the large Yukawa coupling of the top quark to
$H_U$, which
gives the top its mass, also drives the mass-squared parameter
$\mu_U^2$ of $H_U$
negative (with the help of the QCD coupling), while the absence of a
 large
Yukawa in the down sector keeps the mass-squared of $H_D$ positive.
In fact,
conditions (\ref{eq:bndbel}) and (\ref{eq:ewsb}) are easily
satisfied for
a large range of initial conditions if $\lambda_G \sim {\cal O}(1)$,
resulting in a very natural picture of radiative symmetry breaking.
This
picture is essentially lost in the large $\tan\beta$ scenario, for
the following two reasons:
\begin{enumerate}
\item Since both Yukawas are comparable [and in fact initially equal
in the
SO(10) case], the two Higgs doublets tend to run in the same way, so
either
both stay positive at the electroweak scale and the symmetry does
not break, or
both become negative and the potential becomes unbounded from below
(a situation which breaks the symmetry but in a Coleman-Weinberg
fashion,
yielding an ``electroweak scale'' orders of magnitude higher than
the SUSY-breaking scale). The effects which differentiate the
evolution of the two
Higgs doublets, namely hypercharge and the absence of a right-handed
neutrino,
are small and a poor replacement for the usual
$\lambda_t\gg\lambda_b$
splitting. Interestingly, an ${\cal O}(1)$ splitting between
$\lambda_t$ and $\lambda_{b,\tau}$ is still of little use since it
is quickly diminished by the fixed-point behavior of these
couplings. (Some of these observation have been previously made by
T.~Banks \cite{ref:tb}.)
\item Even when electroweak symmetry is broken, a large hierarchy of
VEVs must
be generated between the two similarly-evolving Higgs doublets.
Minimizing the
potential $V_0$ when $\tan\beta \gg 1$ yields

\begin{equation}
-2 m_U^2 = m_Z^2
\label{eq:zmass}
\end{equation}
and
\begin{equation}
{1\over\tan\beta} = -{\mu B\over m_u^2+m_D^2} = -{\mu B\over
m_A^2}\,.
\label{eq:tanbet}
\end{equation}
The first equation sets the scale, but from the second equation we
see that a
large hierarchy in VEVs requires the large hierarchy $\mu B \ll
m_U^2+m_D^2$.
This, as we show below, implies a degree of fine-tuning between some
parameters in the Lagrangian.
\end{enumerate}

\section{Solutions of the RG equations (I)}

Before analyzing the implications of these two criticisms, we
present the
1-loop solutions \cite{ref:us2}\ of the RG equations for the MSSM
mass parameters, integrated
between $M_G = 3\times10^{16}\rm\,GeV$ and a typical squark mass of
300 GeV. (None
of our results is sensitive to the exact values of these starting
and stopping
scales.) The solutions depend on the dimensionless initial values of
the gauge
and Yukawa couplings $\alpha_G$ and $\lambda_G$ and on the
dimensionful
GUT-scale parameters $M_{sq}$, $M_H$, $\mu_G$, $M_{1/2}$, $A_G$,
$B_G$ and
$M_X$. $M_{sq}$ and $M_H$ are the soft-breaking masses of the
$\underline{10}_H$ and the $\underline{16}_3$ respectively, and
$M_X$ will be
explained below; we note for now that it vanishes for universal
soft-breaking masses.  Whenever possible, capital letters denote
values at
the GUT
scale.  The RG equations themselves are well-known and will not be
presented
here. These equations dictate that the
low-energy
values of the various mass parameters depend very simply on the
dimensionful
initial values, with coefficients that depend only on the
dimensionless ones.
Since $\alpha_G$ is known from gauge unification, these coefficients
depend
only on $\lambda_G$. For the representative value $\lambda_G = 1$,
the
solutions are:
\begin{eqnarray}
2\,{m_U^2} &=&
 -5.1\,{{{M_{1/2}^G}}^2} +
 1.2\,{{{M_H^G}}^2} -
 1.6\,{{{M_{sq}^G}}^2} +
 2 \mu^2 -
 3.8\,{{{M_X^G}}^2} \label{eq:mu}\\
{m_A^2} &=&
 -4.9\,{{{M_{1/2}^G}}^2} +
 1.1\,{{{M_H^G}}^2} -
 1.7\,{{{M_{sq}^G}}^2} +
 2 \mu^2 +
 .01\,{{{M_X^G}}^2} \label{eq:ma}\\
{m_Q^2} &=&
 +4.6\,{{{M_{1/2}^G}}^2} -
 .25\,{{{M_H^G}}^2} +
 .51\,{{{M_{sq}^G}}^2} \phantom{\vphantom{x}+ 2\mu^2}+
 1.0\,{{{M_X^G}}^2} \label{eq:mq}\\
{m_t^2} &=&
 +4.1\,{{{M_{1/2}^G}}^2} -
 .27\,{{{M_H^G}}^2} +
 .46\,{{{M_{sq}^G}}^2} \phantom{\vphantom{x}+ 2\mu^2}+
 .85\,{{{M_X^G}}^2} \label{eq:mt}\\
{m_b^2} &=&
 +4.2\,{{{M_{1/2}^G}}^2} -
 .23\,{{{M_H^G}}^2} +
 .55\,{{{M_{sq}^G}}^2} \phantom{\vphantom{x}+ 2\mu^2}-
 2.9\,{{{M_X^G}}^2} \label{eq:mb}\\
{m_L^2} &=&
 +.53\,{{{M_{1/2}^G}}^2} -
 .12\,{{{M_H^G}}^2} +
 .77\,{{{M_{sq}^G}}^2} \phantom{\vphantom{x}+ 2\mu^2}-
 3.1\,{{{M_X^G}}^2} \label{eq:ml}\\
{m_{\tau}^2} &=&
 +.15\,{{{M_{1/2}^G}}^2} -
 .23\,{{{M_H^G}}^2} +
 .55\,{{{M_{sq}^G}}^2} \phantom{\vphantom{x}+ 2\mu^2}+
 1.2\,{{{M_X^G}}^2} \label{eq:mtau}\\
{A_t} &=& +.09\,{A_G} +
 1.8\,{M_{1/2}^G} \label{eq:at}\\
{A_b} &=& +.07\,{A_G} +
 1.9\,{M_{1/2}^G} \label{eq:ab}\\
{A_\tau} &=& +.20\,{A_G} -
 .17\,{M_{1/2}^G} \label{eq:atau}\\
B &=& -
.86\,{A_G}
- 1.1\,{M_{1/2}^G} +
 1.0\,{B_G} \label{eq:bb}\\
\mu  &=& .44\,{\mu_G}
\end{eqnarray}
Here $Q$ and $L$ are the squark and slepton doublets, respectively,
and $t$,
$b$ and $\tau$ are the SU(2)-singlet squarks and sleptons. For
clarity of
presentation, we have dropped from the first 7 expressions above the
terms
proportional to $A_G^2$ and to $A_G M_{1/2}$, since their
coefficients are
small enough [${\cal O}(0.01-0.1)$] and exhibit sufficiently small
custodial-SU(2) breaking to be negligible for purposes of
symmetry-breaking, at least if $A_G$ is
not very much larger than the other mass parameters. We have also
used the
low-energy value of $\mu$ in Eqs.~(\ref{eq:mu}-\ref{eq:ma}). These
solutions
are useful references for the discussions below. These solutions can
also be combined \cite{ref:us2}\ with the more analytic approach
briefly described in Eqs.~(\ref{eq:expsol}) to give a complete,
semi-analytic solution to the RG equations when custodial SU(2) is
an approximate symmetry.

\section{Obtaining a hierarchy of VEVs}

To understand the second criticism above, let us examine in more
detail how Eq.~(\ref{eq:tanbet}) may be satisfied. We
concentrate for the moment on six relevant electroweak-scale
parameters: the up- and down-type Higgs masses $m_U^2$ and $m_D^2$,
a typical squark mass $m_0^2$, the $B$ parameter in the scalar
potential, a gaugino mass (specifically the wino mass) $m_{1/2}$,
and the $\mu$ parameter. This last one may be set to zero by
imposing a Peccei-Quinn symmetry on the Lagrangian, so the size of
$\mu$ measures the breaking of this ${\cal PQ}$ symmetry;
consequently, $\mu$ is multiplicatively renormalized. The previous
two, $B$ and $m_{1/2}$, along with the $A$ parameter, transform in
the same way under a continuous ${\cal R}$ symmetry (so they enter
into
each other's RG equations) and may be made arbitrarily small by
imposing this ${\cal R}$ symmetry. With these two symmetries in mind
\cite{ref:us}, we consider three
possible spectra having splittings which lead to a large $\tan\beta$
according to Eq.~(\ref{eq:tanbet}):

\medskip
\begin{tabular}{||r||c|c|c||}
\hline
mass: & scenario A: & scenario B: & scenario C:
\\ \hline
$7\,m_Z$ &{} &{} &$\pha m_D\,m_0\, \phb\pha\phb$
\\ \hline
$ m_Z$ &$m_U\,m_D\,m_0\,\phb\pha\phb$ &$m_U\,m_D\,m_0\,\phb
m_{1/2}\,\mu\,$
&$m_U\,\pha\pha B\,m_{1/2}\,\mu\,$
\\ \hline
$\vphantom{\textstyle P\over\textstyle p}{1\over7}\,m_Z$
&$\pha\pha\pha B\,m_{1/2}\,\mu\,$ &{} &{}
\\ \hline
$\vphantom{\textstyle P\over\textstyle p}{1\over50}\,m_Z$ &{}
&$\pha\pha\pha B\,\pha\phb$ &{}
\\ \hline
\end{tabular}
\medskip

\noindent Of course there are many other ways to split the
parameters and
obtain the correct hierarchy, but these will suffice to demonstrate
the
fine-tuning involved in the splittings. The value of $\tan\beta$ is
determined
directly only by $m_U$, $m_D$, $\mu$ and $B$; we include $m_0$ and
$m_{1/2}$
to illustrate the symmetries. Scenario A involves no tuning
at all (at this stage of the analysis):
the only hierarchies present are those enforced by the two
symmetries.
However, as also pointed out by Nelson and Randall \cite{ref:nr},
such a scenario is ruled
out for large $\tan\beta$ since it would imply a light chargino, in
disagreement with bounds from LEP \cite{ref:lep}. In fact, both
$\mu$ and $m_{1/2}$ must be
comparable to or above the Z mass to satisfy this bound. Therefore
we {\it
must} widely split some parameters without a symmetry justification,
and this will
entail fine-tuning. For example, in scenario B all parameters are
kept at the
Z mass but the $B$ has been tuned to be light (namely, its initial
value at
the GUT scale is chosen to almost completely cancel the
contributions induced
by $A$ and $m_{1/2}$ through the RG evolution). Alternatively, in
scenario C,
$m_U$ is adjusted to end up much below the other scalar masses
(yielding
$m_A^2 \simeq 50 m_Z^2$) while the other parameters are kept at the
Z mass
using the approximate symmetries. In these two scenarios, and in
fact {\it
generically} whenever $\mu > m_Z$ and $m_{1/2} > m_Z$, the initial
conditions
at the GUT scale must be adjusted to at least a relative accuracy of
$1/\tan\beta$ to obtain the necessary hierarchy of VEVs. We should
point out,
however, that such a tuning is no worse than the one which would be
needed in
the small $\tan\beta$ case if the squarks were experimentally
determined to be
above 700 GeV or so.

\section{Splitting the Higgs doublets}

We return now to the first criticism above, and address the
splitting between
the two Higgs doublets. Recall that after running we need $2 m_U^2 <
0$ while
$m_U^2 + m_D^2 > 0$. However, the two masses evolve almost in
parallel, since custodial symmetry breaking effects, namely
hypercharge and the absence of $\nu_R$, are small.
Thus, if at the GUT scale the mass parameters are custodial-SU(2)
symmetric, the splitting of the two Higgs masses at the weak scale
is small relative to a typical SUSY
mass $M_S$ at the GUT scale: $m_D^2 - m_U^2 \equiv \epsilon_c
M_S^2$ (``$c$'' for custodial). Putting
these together, we learn that only within a window of size $\sim
\epsilon_c$ in the GUT-scale parameter space can we simultaneously
satisfy $m_U^2 < 0$ and $m_A^2 > 0$; if they are satisfied, then
$m_Z^2 = -2 m_U^2 < \epsilon_c M_S^2$ and $m_A^2 = m_U^2 + m_D^2 <
\epsilon_c M_S^2$. In practice, this is usually accomplished
\cite{ref:yukuni}\ using the gaugino mass as the largest mass
parameter, so $M_S = M_{1/2} \geq M_{sq,H}$: this is because,
according to Eqs.~(\ref{eq:mu},\ref{eq:ma}), custodial
breaking effects proportional to $M_{1/2}^2$ lower $m_U^2$ with
respect
to $m_D^2$, while those from the scalar masses $M_{sq,H}^2$ act in
the opposite
way. Furthermore $\mu$ must also typically be ${\cal O}(M_{1/2})$ in
order to keep $m_A^2$ positive.
Then, in addition to the ${\cal O}(\epsilon_c)$
fine-tuning of the Z mass, the large values of $m_{1/2}$ and $\mu$
mean that the $B$ parameter must be adjusted beyond the
${\cal O}(1/\tan\beta)$ accuracy of the previous paragraph. To see
this, we rewrite Eq.~(\ref{eq:tanbet}) in the form
\begin{equation}
{B\over m_{1/2}} = {1\over\tan\beta} {m_U^2 +
m_D^2\over\mu\,m_{1/2}}
\label{eq:btune}
\end{equation}
which quantifies the needed suppression of the electroweak-scale
value of $B$
(achieved by fine-tuning its GUT-scale value) relative to the
minimum value it
would naturally have, namely the value $\sim M_{1/2}$ induced
through the RG
evolution. In the present case, using $\mu \sim M_{1/2} \sim M_S$ we
obtain
$B/m_{1/2} \sim (1/\tan\beta)\, \epsilon_c$.

\section{D-terms}

This highly unnatural state of affairs arises partly because of the
degeneracy of the Higgs doublets and their subsequent parallel
evolution. A possible remedy is actually generic in SO(10)
unification, due to the rank of this group which exceeds by one the
rank of SU(5) or the standard model. Thus we write $\rm SO(10)
\supset SU(5) \otimes U(1)_X$, where $\rm U(1)_X$ is proportional to
$\rm 3(B-L)+4T_{3R}$ [the generator of baryon- minus lepton-number
symmetry and a generator of $\rm SU(2)_R$] and couples to the scalar
fields according to the following table:

\medskip
\begin{tabular}{||r||c|c|c|c|c|c|c||c|c||}
\hline
field:&$H_U$&$H_D$&$Q$&$t$&$b$&$L$&$\tau$&
$\langle\underline{16}_H\rangle$&
$\langle\underline{\overline{16}}_H\rangle$
\\ \hline
$\rm U(1)_X$ charge:&$-2$&$2$&$1$&$1$&$-3$&$-3$&$1$& $5$& $-5$
\\ \hline
\end{tabular}
\medskip

\noindent The $\underline{16}_H$ and $\underline{\overline{16}}_H$
are examples of extra
superheavy Higgs representations which are typically added in order
to break
this $\rm U(1)_X$ (in this case by acquiring VEVs in the ``$\nu_R$''
direction) and reduce the rank of the group. As usual, the
spontaneous
breakdown of a U(1) leads to a VEV for its D-term, which can induce
masses for
the various fields which appear in this D-term. In particular, {\it
if we do
not assume universal soft-breaking masses} for all scalars, then the
soft-breaking masses of the $\underline{16}_H$ and
$\underline{\overline{16}}_H$ need not be equal, and therefore their
VEVs are
also split, in proportion to their mass splitting. This splitting
in turn generates a mass splitting in the low-energy MSSM Lagrangian
through the cross-term:
\begin{equation}
{\cal L} \supset {1\over 2} D_X^2 =
{1\over 2}\left(\langle|\underline{16}_H|^2\rangle -
\langle|\underline{\overline{16}}_H|^2\rangle + 2 |H_U|^2 - 2
|H_D|^2 + \ldots
\right)^2\,.
\end{equation}
In fact, this mechanism splits any fields which have different
charges under
$\rm U(1)_X$. Thus the boundary conditions for the scalar masses at
the GUT
scale become
\begin{eqnarray}
M_{U\phantom{,x,x}}^2 & = & M_H^2 + \mu^2 - 2 M_X^2 \nonumber \\
M_{D\phantom{,x,x}}^2 & = & M_H^2 + \mu^2 + 2 M_X^2
\label{eq:initcond} \\
M_{Q,t,\tau}^2 & = & M_{sq}^2 \phantom{\vphantom{x} + \mu^2} +
\phantom{2}
M_X^2 \nonumber \\
M_{b,L\phantom{,x}}^2 & = & M_{sq}^2 \phantom{\vphantom{x} + \mu^2}
- 3 M_X^2
\nonumber
\end{eqnarray}
where the capital letters on the left-hand side serve as reminders
that these
are the values at the GUT scale, and
\begin{equation}
M_X^2 = {1\over 10} (M_{16}^2 - M_{\overline{16}}^2)
\label{eq:mx}
\end{equation}
is a new soft-breaking mass parameter in the low-energy theory. With
this mass we no longer need rely on large gaugino masses to split
the Higgs
doublets: they can start out being different, and thus even with
parallel
evolution the symmetry-breaking conditions
(\ref{eq:bndbel}--\ref{eq:ewsb}) can apparently be satisfied.

One problem with this mechanism is evident from the initial
conditions in
Eq.~(\ref{eq:initcond}): not just the Higgs doublets but also the
squarks and
sleptons are split, so an excessively large $M_X^2$ could lower
$M_b^2$ or
$M_L^2$ sufficiently to make $m_b^2$ or $m_L^2$ negative at the
electroweak
scale, thereby spontaneously breaking the strong or electromagnetic
gauge
symmetries. If RG effects were irrelevant, namely for small
$\lambda_G$, then
$M_{sq}$ could always be raised enough to prevent this without
affecting $m_{U,D}^2$. However, for
$\lambda_G \sim {\cal O}(1)$ the squark masses strongly affect the
evolution
of the Higgs doublets [see Eqs.~(\ref{eq:mu}-\ref{eq:ma})], and only
for very constrained ranges of the initial parameters can the
electroweak symmetry, and only that symmetry, be spontaneously
broken at a reasonable scale. In fact, as we show in brief below,
there is a focusing effect that is inherent in the MSSM RG equations
when $\lambda_b \sim \lambda_t$, and which inevitably requires an
adjustment of the GUT-scale parameters beyond the $1/\tan\beta$
level derived above. We first show this behavior of the RG equations
for completely general initial conditions, and then return to
discuss the specific case of Eqs.~(\ref{eq:initcond}).

\section{Solutions of the RG equations (II)}

Consider the RG equations of the MSSM in the limit of exact ${\cal
PQ}$ and ${\cal R}$ symmetries, in which $\mu = M_{1/2} = A = B = 0$
at all scales. For future reference, we call this scenario {\it the
maximally symmetric case}. This limit is interesting for two
reasons: First, no large corrections arise to the $b$ quark mass,
and the $R=m_b/m_\tau$ prediction for all values of $\lambda_G \sim
{\cal O}(1)$ falls nicely within the range allowed by experiment
(see Fig.~1); in other words, a heavy top quark near its fixed-point
mass favors small $\delta m_b$. Second, as we saw above, having a
large $\mu$ and $m_{1/2}$ calls for fine-tuning $B$ (or some
equivalent adjustment), so we'd like to explore the opposite limit
to see whether a more natural scenario can be achieved. Of course,
eventually we must relax this limit to agree with LEP bounds, but
the qualitative behavior we shall discover will persist. If we
further  approximate $\lambda_b \simeq \lambda_t \equiv \lambda$ and
neglect the sleptonic contributions (thereby restoring custodial
symmetry), the RG solutions simplify considerably. There are now
five relevant parameters. In terms of their initial conditions at
the GUT scale, $M_U^2$, $M_D^2$, $M_Q^2$, $M_t^2$ and $M_b^2$, the
solutions at the electroweak scale are:
\begin{eqnarray}
-2 m_U^2 & = & -\fra37 \epsilon_\lambda X
               -\fra35 \epsilon'_\lambda X'
               - \phantom{\fra14} I
               - \phantom{\fra14} I' \nonumber \\
\phantom{-2}m_A^2    & = & \phantom{-}\fra37 \epsilon_\lambda X
               \phantom{\vphantom{x}-\fra35 \epsilon'_\lambda X'}
               + \phantom{\fra14} I
               - \phantom{\fra14} I' \nonumber \\
\phantom{-2}m_Q^2    & = & \phantom{-}\fra17 \epsilon_\lambda X
               \phantom{\vphantom{x}-\fra35 \epsilon'_\lambda X'}
               - \fra14 I
               \phantom{\vphantom{x}-\fra14 I'}
               + \fra14 I'' \label{eq:expsol} \\
\phantom{-2}m_t^2    & = & \phantom{-}\fra17 \epsilon_\lambda X
               +\fra15 \epsilon'_\lambda X'
               - \fra14 I
               -\fra12 I'
               - \fra14 I'' \nonumber \\
\phantom{-2}m_b^2    & = & \phantom{-}\fra17 \epsilon_\lambda X
               -\fra15 \epsilon'_\lambda X'
               - \fra14 I
               +\fra12 I'
               - \fra14 I'' \nonumber
\end{eqnarray}
where
\begin{eqnarray}
\epsilon_\lambda &=&
\exp\left(-{7\over8}\!\!\!\!\!\!\int\limits_{\phantom{xxx}\ln
m_Z}^{\phantom{xxx}\ln M_G}{ {\lambda^2\over\pi^2}
\,d\ln\mu}\right) \label{eq:epsl}\\
&\sim&0.085 \qquad({\rm for}\,\lambda_G \simeq 1)\,,\nonumber \\
\epsilon'_\lambda &=& \epsilon_\lambda^{5/7}\,,
\label{eq:epslp}
\end{eqnarray}
and
\begin{eqnarray}
X\phantom{^\prime} &=& M_U^2 + M_D^2 + 2 M_Q^2 + M_t^2 + M_b^2
\nonumber \\
X' &=& M_U^2 - M_D^2 \phantom{\vphantom{}+ 2 M_Q^2} + M_t^2 - M_b^2
\nonumber \\
I\phantom{^\prime} &=& \fra47 (M_U^2 + M_D^2) - \fra37 (2 M_Q^2 +
M_t^2 + M_b^2)
\label{eq:xis} \\
I' &=& \fra25 (M_U^2 - M_D^2) - \fra35 (M_t^2 - M_b^2)
\nonumber \\
I'' &=& \phantom{\fra25 (M_U^2 - M_D^2)} 2 M_Q^2 - M_t^2 - M_b^2\,.
\nonumber
\end{eqnarray}
(Note again the use of capital letters to denote GUT-scale initial
parameters, and recall that $U$, $D$, $Q$, $t$ and $b$ refer to the
up-type Higgs, the down-type Higgs, the SU(2)-doublet
third-generation squarks, the SU(2)-singlet stop and the
SU(2)-singlet sbottom, respectively.)

Evidently, two linear
combinations of masses, labeled by $X$ and $X'$ at the GUT scale,
renormalize multiplicatively and exponentially contract at low
energies for $\lambda_G \sim {\cal O}(1)$. The three other linear
combinations, $I$, $I'$ and $I''$, are invariant.  The important
observation here is that in the first contraction---the sum rule
$m_A^2 + 2 m_Q^2 + m_t^2 + m_b^2 = \epsilon_\lambda
X$---the coefficient of every term is positive, while we already
know
that each mass-squared itself must be positive for a proper
electroweak-breaking scenario. Therefore each term by itself must be
small, less than $\epsilon_\lambda X$.  This can only happen if the
various combinations of invariants (and possibly also
$\epsilon'_\lambda X'$) in the expressions (\ref{eq:expsol}) for
these terms are adjusted to be small relative to $X$. The exact
constraints that follow from this requirement are given explicitly
elsewhere \cite{ref:us2}. They are of the form $I,\,I',\,I''
\roughly{<} \max(\epsilon_\lambda X,\epsilon'_\lambda X')$.  We
learn that, for $\lambda_G \sim {\cal O}(1)$ where this focusing
effect is important, any given model for the GUT-scale soft-breaking
masses must either provide an explanation of why each invariant
should be small relative to the sum $X =  M_U^2 + M_D^2 + 2 M_Q^2 +
M_t^2 + M_b^2$, or else that invariant must be tuned by hand to be
small.  We also learn that the conditions for successful
symmetry-breaking are sensitive to any other small effects. One such
effect is custodial symmetry violation, which is parametrized above
by $\epsilon_c$ and results from hypercharge and $\lambda_\tau$ (or
the absence of $\nu_R$). In the running of the Yukawas, both of
these cause $\lambda_t$ to slightly exceed $\lambda_b$, and thus
drive $m_U^2$ below $m_D^2$, as in the conventional scenario of
electroweak symmetry breaking.
In the running of the masses, the contributions of the $\tau$ Yukawa
have an opposite and numerically more relevant impact. Therefore the
custodial-breaking effects make it harder to break the symmetry
correctly---significantly harder in the specialized case discussed
below. In any realistic scenario there are also contributions from
the gauginos
and $\mu$, so in the end the sum rule, and therefore the general
limit which
must be set on $I$, $I'$ and $I''$, takes the form
\begin{equation}
\left\{I,I',I'',m_A^2 + 2 m_Q^2 + m_t^2 + m_b^2\right\} \roughly{<}
{\cal
O}\left[\max\left(\epsilon_\lambda,\epsilon_c,
{\mu^2\over M_S^2},{m_{1/2}^2\over M_S^2}\right)\right] M_S^2
\label{eq:sumrul}
\end{equation}
where $M_S$ is the largest mass parameter in the initial conditions
at the GUT scale.

If we now return to the more specialized boundary conditions of
Eq.~(\ref{eq:initcond}), we find (after setting $\mu = 0$):
\begin{eqnarray}
X\phantom{^\prime} &=& 2 M_H^2 + 4 M_{sq}^2 \nonumber \\
X' &=& 0 \nonumber \\
I\phantom{^\prime} &=& \fra47 (2 M_H^2 - 3
M_{sq}^2)\label{eq:spexis} \\
I'&=& -4 M_X^2 \nonumber \\
I''&=& +4 M_X^2\,. \nonumber
\end{eqnarray}
For this choice of boundary conditions, keeping the invariants small
imposes only two requirements:
\begin{equation}
M_H^2 - \fra32 M_{sq}^2 = \epsilon_{\lambda c} M_S^2
\ll M_S^2 \label{eq:mhmsq}
\end{equation}
and small $M_X$. The upper bound on the invariants involves both
$\epsilon_\lambda$ and $\epsilon_c$, and as we hinted above they
partially cancel in the combination $\epsilon_{\lambda c}$ which
enters into the requirements: $|\epsilon_{\lambda c}| <
|\epsilon_\lambda|,|\epsilon_c|$. (One sum rule which can be formed
under these boundary conditions, $-2 m_U^2 + \fra43 m_A^2 + \fra43
m_b^2$, is particularly sensitive to this
cancellation\cite{ref:us2}.) The first requirement,
Eq.~(\ref{eq:mhmsq}),  entails a definite tuning of parameters to a
precision of $\epsilon_{\lambda c}$, which does not apparently
follow from any symmetry.  Note that without this requirement,
either color breaks
when $M_H^2 > \fra32 M_{sq}^2$ or a Coleman-Weinberg mechanism
operates when $M_H^2 < \fra32 M_{sq}^2$.
The requirement of small $M_X$ may on the other hand
be natural, since [see Eq.~(\ref{eq:mx})] the value of $M_X^2$ is
smaller by an order of magnitude than the soft-breaking masses whose
splitting generates the D-terms, and those masses may be expected to
be comparable to $M_{sq}$ and $M_H$.
In any case, we see that
because of the focusing effect of the RG equations, the D-terms
cannot be
{\it allowed} to induce splittings bigger than those we already had
through
custodial SU(2)-breaking effects. Hence they do not eliminate the
criticism that the electroweak symmetry is hard to break when the
Yukawas are comparable. But there is still a significant advantage
in
using these D-terms, since they can now substitute for large values
of
$m_{1/2}$ and $\mu$, and with light gauginos and $\mu$ it is much
easier to
obtain a large $\tan\beta$, according to Eqs.~(\ref{eq:tanbet}) and
(\ref{eq:btune}).

\section{Radiative bottom decay (II)}

Before putting the various observations to use in examining specific
scenarios and their merits, we point out another feature of the
solutions
to the RG equations which will further constrain the scenarios. As
is evident from Eqs.~(\ref{eq:at}-\ref{eq:atau}) (or directly from
the RG
equations), the initial value $A_G$ hardly affects the low-energy
values of
$A_{t,b,\tau}$; they are instead largely determined in magnitude and
sign by
the gaugino mass $M_{1/2}^G$, which also fixes the low-energy gluino
mass. (It
is difficult, though perhaps not impossible for sufficiently small
$\lambda_G$, to construct models in
which $A_G
\gg M_{1/2}$ and yet the electroweak symmetry but neither color nor
charge
breaks spontaneously and correctly, so we shall disregard this
possibility in these proceedings. The implications of tuning $A_G$
to cancel the gaugino mass at low energies in the expression for
$A_t$ will be considered elsewhere \cite{ref:us2}.)
This observation, which was also emphasized by Carena et
al.~\cite{ref:copw}, directly relates
the $\delta m_b$ corrections of Eq.~(\ref{eq:delmb}) to the large
$b\rightarrow s\gamma$ graphs discussed above. (More precisely, the
gluino-
and higgsino-exchange diagrams for each process are directly
related.) The
sign of this correlation \cite{ref:copw}\  is such
that when $\delta m_b < 0$ (i.e. the predicted $R$ is lowered, and
therefore
so is the top mass) then the large $b\rightarrow s\gamma$ graphs
interfere
{\it constructively} with the usual 2-Higgs standard model
amplitude, and
vice-versa.  On one hand, we see from Eq.~(\ref{eq:muglimits}) or
from Fig.~1, that the  bounds on $\delta m_b$ are more severe when
$\delta m_b > 0$.
On the other hand,
as noted above, when $\delta m_b < 0$ the interference is
constructive and the bounds on the large
$b\rightarrow s\gamma$ graphs are stricter. Thus these two bounds
are much
stronger when taken together, and translate into the following
statement:
either (a) the gauginos or $\mu$ or both are significantly lighter
than the
squarks, or (b) the superpartners are much heavier than the Z.

\section{Case studies}

With these remarks in mind, we first examine the popular
\cite{ref:yukuni}\ case of universal
soft-breaking masses. This scenario has also been recently studied
in some
detail by Carena et al.~\cite{ref:copw}. If all soft-breaking scalar
masses are equal then the
D-term contributions vanish ($M_X \equiv 0$), and we are left with
the three
parameters $\mu$, $M_{1/2}$ and $M_0\equiv M_{sq} = M_H$, in
addition to $B_G$
which is adjusted at the end to obtain the correct $\tan\beta$ [see
Eq.~(\ref{eq:tanbet})]. We have already mentioned that in this case
we need $\mu$ comparable to $M_{1/2}$, and both at least as big as
$M_0$, to break electroweak symmetry correctly. We have also seen
that these three parameters must be tuned in order to obtain a
positive $m_Z^2$ and $m_A^2$:
\begin{equation}
m_Z^2 \sim m_A^2 \sim \epsilon_c M_{1/2}^2\,.
\label{eq:ufta}
\end{equation}
Next, to achieve a hierarchy of Higgs VEVs, $B_G$ must be adjusted
very
precisely such that, at low energies,
\begin{equation}
{B\over m_{1/2}} = {1\over\tan\beta} {m_A^2\over\mu\,m_{1/2}}
\sim {\epsilon_c\over\tan\beta}\,.
\label{eq:uftb}
\end{equation}
Finally, since $\mu$ and the gauginos are not lighter than the
squarks,
$\delta m_b$ is rather large, and so must be negative, as can be
seen from
Eq.~(\ref{eq:muglimits}) or from Fig.~1. Hence the $b\rightarrow
s\gamma$
constraint is strong, necessitating large superpartner masses of at
least
${\cal O}(\rm TeV)$ and therefore a {\it further} tuning (by roughly
another order
of magnitude) of the three parameters to achieve correct electroweak
breaking.
Of course, such a highly-tuned scenario is also highly predictive:
for
example, the spectrum is highly constrained, and the top mass is
predicted by
the large $\delta m_b$ corrections to be light.

In search of a more natural scenario, we next relax the assumption
of
universal soft-breaking masses, which was perhaps arbitrary to begin
with. We begin with the maximally symmetric scenario studied
previously (see also scenario C above), in which
$\mu\sim M_{1/2}\sim A_G\sim B_G\,(\sim m_Z) \ll {m_A,\ldots,m_b}
\,(\ll M_{sq}\sim M_H)$ so
that
the ${\cal PQ}$ and ${\cal R}$ symmetries are approximately obeyed.
Since this
hierarchy directly implies a small $\delta m_b$ and therefore (from
Fig.~1) a
relatively large $\lambda_G$, the focusing effect of
Eq.~(\ref{eq:epsl}) takes
its toll, and once again the initial parameters---this time
$M_{sq}$, $M_H$,
and $M_X$---must be adjusted to at least ${\cal
O}\left[\max\left(\epsilon_\lambda,\epsilon_c\right)\right]$. Then,
to truly get a maximally symmetric scenario, we choose to obtain a
large $\tan\beta$ not by tuning to get a small $B$ but rather by
(equivalently) tuning to get a small $m_Z^2 \sim m_A^2/\tan\beta$,
which then allows small $\mu$ and $m_{1/2}$ relative to the typical
low-energy soft-breaking masses and therefore establishes
approximate ${\cal PQ}$ and ${\cal R}$ symmetries even at the
electroweak scale:
\begin{equation}
-2 m_U^2 = m_Z^2 \sim
{\max\left(\epsilon_\lambda,\epsilon_c\right)\over\tan\beta}
M_S^2
\label{eq:sfta}
\end{equation}
where $M_S \equiv M_H \sim M_{sq}$.
After this adjustment, large $\tan\beta$ is automatic:
\begin{equation}
{B\over m_{1/2}} = {1\over\tan\beta} {m_A^2\over\mu\,m_{1/2}}
\sim {\cal O}(1)\,.
\label{eq:sftb}
\end{equation}
And no further adjustment is necessary to suppress $b\rightarrow
s\gamma$ since
$\mu$ and $M_{1/2}$ are small. So this scenario requires much less
adjusting
than the universal case, although more than just the inevitable
$1/\tan\beta$
tuning. (The actual tuning needed is in reality slightly more than
indicated, due to the squark- and slepton-splittings induced by the
D-terms and due to the effects of custodial symmetry violation, as
mentioned above; qualitatively, though, the picture we described
remains.) It is also predictive: the superspectrum is hierarchical
with light
charginos and neutralinos but heavy squarks, and since $\lambda_G$
is large so is the top mass.

We can continuously retreat from this maximally symmetric case by
increasing $\mu$ or $M_{1/2}$ (or the related parameters), thereby
losing ${\cal PQ}$ or ${\cal R}$, first at low energies when $\mu$
or $m_{1/2}$ become comparable to $\epsilon_\lambda^{1/2} M_S$, and
then at all energies when $\mu$ or $m_{1/2}$ become comparable to
$M_S$ itself. Both may be interesting for model-building (for
example, if $\mu$ is radiatively generated by $A$ terms at the GUT
scale, then ${\cal PQ}$ rather than ${\cal R}$ symmetry should be
evident) and for comparison with experiments once the superspectrum
is measured. In either case, the tuning is comparable to the
maximally symmetric case, though less tuning is needed in
Eq.~(\ref{eq:sfta}) and more in Eq.~(\ref{eq:sftb}). We will however
defer their discussion to our more complete study \cite{ref:us2},
and instead consider the case where {\it both} ${\cal PQ}$ and
${\cal R}$ symmetries are abandoned in favor of a smaller
$\lambda_G$.
We will call this scenario, unimaginatively, the
asymmetric case. It alleviates the
focusing effect of Eq.~(\ref{eq:epsl}) since now $\epsilon_\lambda
\sim {\cal
O}(1)$, which in turn allows the initial conditions (specifically
the D-terms) to split the Higgs doublets
and the other multiplets by a large amount, so now $M_X \sim M_S$.
Small $\lambda_G$ also entails larger (negative) $\delta m_b$
corrections to correctly predict $m_b/m_\tau$, and to this end we
take $\mu\sim M_{1/2} \sim M_{sq} \sim M_H \equiv M_S$. We then see,
however, that what we
gain by eliminating the focusing effect we lose by restoring the
$b\rightarrow
s\gamma$ problem: since $\mu$ and $m_{1/2}$ are no longer small, we
are forced
to raise the SUSY scale to $\sim {\cal O}(\rm TeV)$, and therefore
again to tune the initial
parameters to make the Z light:
\begin{equation}
-2 m_U^2 = m_Z^2 \sim \left({1\over10} M_S\right)^2.
\label{eq:afta}
\end{equation}
Since the D-term splitting of the Higgs is now large, $m_A^2 \sim
M_S^2$, we find that $B_G$ requires only the typical tuning
\begin{equation}
{B\over m_{1/2}} = {1\over\tan\beta} {m_A^2\over\mu\,m_{1/2}}
\sim {1\over\tan\beta}\,.
\label{eq:aftb}
\end{equation}
This scenario is comparable in its naturalness (or lack thereof) to
the
symmetric case. The superspectrum is uniformly heavy rather than
hierarchical, and the top is light since $\lambda_G$ is small.

\section{Conclusions}

These last two scenarios are qualitatively the best one can hope for
from an
SO(10) model with Yukawa unification, for two reasons. First,
obtaining a large $\tan\beta$
and thereby the top-bottom hierarchy is never natural in the MSSM,
due to
the LEP bounds on $\mu$ and $m_{1/2}$ which force either $m_A^2$ to
be much
heavier than the Z or $B$ to be much lighter than the gauginos.
Second, the last two scenarios illustrate how, for large
$\lambda_G$, the
{\it inherent} focusing property of the RG equations in the
symmetric limit necessitates a further tuning of the
initial parameters, while for small $\lambda_G$ a similar tuning is
mandated by bounds on the rate of $b\rightarrow s\gamma$. There are
also intermediate scenarios, with only one of the symmetries, but
they are apparently no more natural. These various possibilities are
considered in more detail elsewhere \cite{ref:us2}. The universal
case is generically much more tuned than most of these scenarios, as
we have shown. Hence departures from universality, although possibly
dangerous for the
flavor-changing neutral current interactions they can
induce in some models, are
strongly favored in achieving a large $\tan\beta$.

What is the status of the predictions? Perhaps surprisingly,
the top mass is not
an independent prediction of Yukawa unification, but rather depends
strongly
(i.e. non-logarithmically) on a certain ratio of superpartner
masses appearing in $\delta m_b$. Since
the large- and small-$\lambda_G$ cases are equally fine-tuned,
naturalness
arguments do not single out any particular top mass within this
SO(10) framework.
Instead, information about the top mass can be combined with
Fig.~1 and
Eq.~(\ref{eq:delmb}) to restrict the mass parameters which can
be
consistent with Yukawa unification, and to determine a favored
superspectrum.
It must be admitted that, with Yukawa unification, the attractive
conventional
picture of {\it radiative} electroweak symmetry breaking due to a
large
$\lambda_t$ but small $\lambda_b$ is largely lost: the symmetry is
either broken radiatively using small custodial isospin-violating
effects and extraordinarily fine-tuned initial conditions, or else
the Higgs doublets are already split at the GUT scale. Furthermore,
we have
seen that all the large $\tan\beta$ scenarios are technically
unnatural. On
the other hand (and to some extent because of the necessary
fine-tuning), they
are certainly predictive and so will be tested in future
accelerators: for
example, if the charginos are light but the squarks
are heavy,
the top should also be heavy; if the $\rm SU(2)_L$-singlet bottom
squark or
the doublet sleptons were lighter than the other squarks and
sleptons, this
would be a sign that the D-term splittings were large
\cite{ref:hitoshi};
and
finally, $\tan\beta$ can itself eventually be measured to decisively
confirm or dismiss the
large $\tan\beta$ hypothesis.

Even before any further experimental input, there are a few
theoretical
avenues worth pursuing which would make Yukawa unification much more
attractive. First, as discussed also by J.~Lykken
in these proceedings \cite{ref:lyk}, string models which lead to
true grand-unified models as
their low-energy effective Lagrangians can exhibit higher symmetries
in
certain sectors of the GUT than in other sectors. In particular, a
string
theory with an SO(10) gauge symmetry might break to an effective
low-energy
SU(5) GUT in a stringy way, leaving {\it all three} Yukawa couplings
unified
at the Planck scale \`{a} la SO(10) but splitting the
$\underline{5}_H$ from
the $\underline{\overline{5}}_H$ soft masses, namely $M_U^2$ from
$M_D^2$. Note that unlike the D-term splittings we have considered
before, $M_U^2$ and $M_D^2$ could now be split without necessarily
decreasing some squark or slepton masses. This approach can thus
provide more freedom in the choice of boundary conditions---although
as we have shown, some features of the RG equations, in particular
the focusing effect, are inherent in the equations themselves in
certain limits and apply to any boundary conditions, so that freedom
could either be
the key to a more natural scenario (see the following remark) or
could necessitate still more arbitrary fine-tunings at the GUT
scale.
Second, we have shown that in the symmetric case with large
$\lambda_G$, the
ratio of the
soft-breaking masses for the $\underline{10}_H$ and the
$\underline{16}_3$
needs to have a certain value if the electroweak symmetry is to
break
correctly. While such a value is arbitrary in the context of an
SO(10) model
and therefore apparently requires a fine-tuning, perhaps this value
could be
explained as a ratio of integral conformal weights in the context of
a string
theory into which the GUT is embedded. (Notice that this value is
favored simply by the unification of Yukawas at some large scale,
and does not depend on an SO(10) symmetry.) If such an explanation
could be found, then the symmetric large-$\lambda_G$ case would now
be strongly
favored: it would only require the
minimal $1/\tan\beta$ tuning (and would predict a heavy top!). In
fact, if the
squarks were to be {\it experimentally determined} to be heavy while
the
gauginos were light, then this case would be no more fine-tuned than
the
conventional small $\tan\beta$ scenario, but would have the
advantage of
explaining the top-bottom mass hierarchy (through the ${\cal PQ}$
and ${\cal R}$ symmetries)---which, after all, was historically the
motivation for studying Yukawa unification.


\section*{Acknowledgments}
We would like to thank M.~Carena and C.~Wagner for enjoyable
discussions, in particular regarding the sign of the correlation
between the diagrams in $b\rightarrow s\gamma$ and those in $\delta
m_b$. We have also benefited from discussions with J.~Lykken and
F.~Zwirner regarding string models. U.S.~would like to thank the
organizers of the Second IFT Workshop on Yukawa Couplings and the
Origins of Mass for a stimulating conference. This work was
supported in part by NSF grants PHY-89-17438 (U.S.), PHY-91-21039
(R.R.) and PHY-90-21139 (L.J.H.) and in part by the United States
Department of Energy under contract no.~DE-AC03-76SF00098 (L.J.H.).
\section*{Figure Captions}
\begin{description}
\item[Fig.~1:] \figone
\item[Fig.~2:] \figtwo
\item[Fig.~3:] \figthree
\end{description}

\end{document}